\documentclass{emulateapj}
\usepackage{apjfonts}
\usepackage{graphicx}

\shortauthors{Winn et al.\ 2011}
\shorttitle{RM effect for Kepler-16}

\begin{document}

%
\def\ltsima{$\; \buildrel < \over \sim \;$}
\def\lsim{\lower.5ex\hbox{\ltsima}}
\def\gtsima{$\; \buildrel > \over \sim \;$}
\def\gsim{\lower.5ex\hbox{\gtsima}}
                                                                                          
%

\bibliographystyle{apj}

\title{ Spin-Orbit Alignment for the Circumbinary Planet Host Kepler-16~A }

\author{
Joshua N.\ Winn\altaffilmark{1},
Simon Albrecht\altaffilmark{1},
John Asher Johnson\altaffilmark{2},
Guillermo Torres\altaffilmark{3},
William D.\ Cochran\altaffilmark{4},\\
Geoffrey W.\ Marcy\altaffilmark{5},
Andrew W.\ Howard\altaffilmark{5},
Howard Isaacson\altaffilmark{5},
Debra Fischer\altaffilmark{6},
Laurance Doyle\altaffilmark{7},
William Welsh\altaffilmark{8},\\
Joshua A.\ Carter\altaffilmark{3},
Daniel C.\ Fabrycky\altaffilmark{9},
Darin Ragozzine\altaffilmark{3},
Samuel N.\ Quinn\altaffilmark{3},
Avi Shporer\altaffilmark{10},
Steve B.\ Howell\altaffilmark{11}, \\
David W.\ Latham\altaffilmark{3},
Jerome Orosz\altaffilmark{8},
Andrej Prsa\altaffilmark{12},
Robert W.\ Slawson\altaffilmark{7},
William J.\ Borucki\altaffilmark{13},
David Koch\altaffilmark{13}, \\
Thomas Barclay\altaffilmark{14},
Alan P.\ Boss\altaffilmark{15},
J{\o}rgen Christensen-Dalsgaard\altaffilmark{16},
Forrest R.\ Girouard\altaffilmark{13,17},
Jon Jenkins\altaffilmark{7,13}, \\
Todd C.\ Klaus\altaffilmark{17},
S{\o}ren Meibom\altaffilmark{3},
Robert L.\ Morris\altaffilmark{7,13},
Dimitar Sasselov\altaffilmark{3},
Martin Still\altaffilmark{14}, 
Jeffrey Van Cleve\altaffilmark{7,13}
}

\journalinfo{Accepted version}
\slugcomment{{\it The Astrophysical~Journal (Letters)}, in press}

 \altaffiltext{1}{Department of Physics, and Kavli Institute for
   Astrophysics and Space Research, Massachusetts Institute of
   Technology, Cambridge, MA 02139}

 \altaffiltext{2}{Department of Astrophysics, California Institute of
   Technology, MC249-17, Pasadena, CA 91125; and NASA Exoplanet
   Science Institute (NExScI)}

 \altaffiltext{3}{Harvard-Smithsonian Center for Astrophysics, 60
   Garden St., Cambridge, MA 02138}

 \altaffiltext{4}{McDonald Observatory, The University of Texas,
   Austin, TX 78712}

 \altaffiltext{5}{Department of Astronomy, University of California,
   Berkeley, CA 94720}

 \altaffiltext{6}{Department of Astronomy, Yale University, New Haven,
   CT 06511}

 \altaffiltext{7}{Carl Sagan Center for the Study of Life in the
   Universe, SETI Institute, 189 Bernardo Ave., Mountain View, CA
   94043}

 \altaffiltext{8}{Astronomy Department, San Diego State University,
  5500 Campanile Drive, San Diego, CA 92182}

 \altaffiltext{9}{Hubble Fellow, Department of Astronomy and
  Astrophysics, University of California, Santa Cruz, CA 95064}

 \altaffiltext{10}{Las Cumbres Observatory Global Telescope Network,
   6740 Cortona Drive, Suite 102, Santa Barbara, CA 93117}

 \altaffiltext{11}{National Optical Astronomy Observatory, Tucson, AZ
   85726}

 \altaffiltext{12}{Department of Astronomy and Astrophysics, Villanova
   University, 800 E.\ Lancaster Ave., Villanova, PA 19085}

 \altaffiltext{13}{NASA Ames Research Center, Moffett Field, CA 94035}

 \altaffiltext{14}{Bay Area Environmental Research Institute/NASA Ames
   Research Center, Moffett Field, CA 94035}

 \altaffiltext{15}{Department of Terrestrial Magnetism, Carnegie
   Institution of Washington, 5241 Broad Branch Road, NW, Washington,
   DC 20015}

 \altaffiltext{16}{Danish AsteroSeismology Centre, and Department of
   Physics and Astronomy, Aarhus University, Ny Munkegade, DK-8000
   Aarhus C, Denmark}

 \altaffiltext{17}{Orbital Sciences Corporation/NASA Ames Research
   Center, Moffett Field, CA 94035}

\begin{abstract}

  Kepler-16 is an eccentric low-mass eclipsing binary with a
  circumbinary transiting planet. Here we investigate the angular
  momentum of the primary star, based on {\it Kepler} photometry and
  Keck spectroscopy. The primary star's rotation period is $35.1\pm
  1.0$~days, and its projected obliquity with respect to the stellar
  binary orbit is $1.6\pm 2.4$~degrees. Therefore the three largest
  sources of angular momentum---the stellar orbit, the planetary
  orbit, and the primary's rotation---are all closely aligned. This
  finding supports a formation scenario involving accretion from a
  single disk. Alternatively, tides may have realigned the stars
  despite their relatively wide separation (0.2~AU), a hypothesis that
  is supported by the agreement between the measured rotation period
  and the ``pseudosynchronous'' period of tidal evolution theory. The
  rotation period, chromospheric activity level, and fractional light
  variations suggest a main-sequence age of 2-4~Gyr. Evolutionary
  models of low-mass stars can match the observed masses and radii of
  the primary and secondary stars to within about 3\%.

\end{abstract}


\section{Introduction}
\label{sec:introduction}

Kepler-16~(AB)-b is a planet with two parent stars (Doyle et
al.~2011). The stars (0.7 and 0.2~$M_\odot$) are in 41-day eccentric
orbit, and the planet (0.3~$M_{\rm Jup}$) circles both of them every
229 days. Viewed from the Solar system, the stars eclipse each other
and the planet transits both of them, providing definitive evidence
that circumbinary planets exist and permitting precise determinations
of the system's parameters. For example the planet's radius is known
to within 0.3\%, better than that of any other known exoplanet. The
stars are themselves of interest as a rare example of low-mass dwarfs
with precisely known dimensions.

Such a unique system should be studied in every possible way, for
exploratory purposes as well as the specific purpose of understanding
its formation and evolution. How old are the stars? Did the planet
form together with the stars, or was it captured from another system?
Has there been tidal evolution or other effects that have modified the
system's architecture? Here we present an investigation of the angular
momentum of the primary star, bearing on these questions.

It has already been established that the planes of the circumbinary
orbit and the stellar orbit are aligned to within $0\fdg5$ (Doyle et
al.~2011). This suggests all three bodies inherited their angular
momentum from a single disk, as opposed to dynamical scenarios that
are often invoked for triple systems such as close encounters (Mikkola
1984, Bailyn 1989, Ivanova 2008) or dynamical decay (Sterzik \&
Tokovinin 2002). One must remember, though, that the planet was
discovered with transit photometry, a technique that is severely
biased toward finding coplanar orbits. This raises the question of
whether the orbital coplanarity of Kepler-16 is at all representative
of circumbinary planets, and motivates measurements of the alignment
between the orbital axes and the stellar spin axes, for which there
was no selection bias.

This {\it Letter} is organized as follows. Section~2 presents a
photometric determination of the rotation period. Section~3 presents a
spectroscopic determination of the sky-projected stellar obliquity
(the angle between the rotation axis of the primary star, and the
stellar orbital axis), based on observations of the
Rossiter-McLaughlin effect. Section~4 discusses the implications of
these results for our understanding of the primary star and of the
system's history.

\vspace{0.5in}

\section{The rotation period}
\label{sec:rotper}

We measured the rotation period of the primary star with data from the
{\it Kepler} spacecraft, a 0.95m space telescope that monitors the
optical brightness of about 150,000 stars in a quest to detect
transits of potentially habitable Earth-sized planets (Borucki et
al.~2010). Overviews of the mission design, the instrument
performance, and the data processing pipeline were given by Koch et
al.~(2010), Caldwell et al.~(2010) and Jenkins et al.~(2010).

Kepler-16 was observed with 29.4~min sampling for a nearly continuous
600-day interval, from 2009~May~02 to 2010~December~22 (quarters 1-7).
The duty cycle was 94\%, with 17 short gaps due to technical problems
as well as scheduled interruptions in observing. After each
interruption a jump was observed in the relative flux. We placed all
the data onto a common flux scale under the assumption that the flux
variations during the interruptions were smooth enough to be described
by a quadratic function of time. Specifically, we multiplied the data
from each of the 18 disjoint intervals by a constant, and determined
the optimal values of the constants by fitting quadratic functions to
the data within one day of an interruption.

The resulting time series exhibited a secular 3\% decrease in relative
flux, which could be an instrumental effect or a true decrease in
stellar brightness. Since this trend is irrelevant to the rotation
period determination, we applied a 70-day median filter prior to
plotting the time series in the top panel of Figure~1. The time series
exhibits quasiperiodic variations of order 0.5\%. As usual for
late-type stars, we attribute these variations to dark spots and
bright plages being carried around by stellar rotation.

The bottom panel of Figure~1 is a Lomb-Scargle periodogram, showing a
prominent peak at 35.1~days along with smaller peaks at the first two
harmonics. We identified this peak with the stellar rotation period.
We estimated the uncertainty in the period by dividing the data
chronologically into 4 equal segments, analyzing each piece
separately, and finding the standard deviation in the mean of the
periodogram peaks. Based on this analysis we find $P_{\rm rot} =
35.1\pm 1.0$~days.

\begin{figure*}[ht]

 \begin{center}
 \leavevmode
 \hbox{
 \epsfxsize=6.5in
 \epsffile{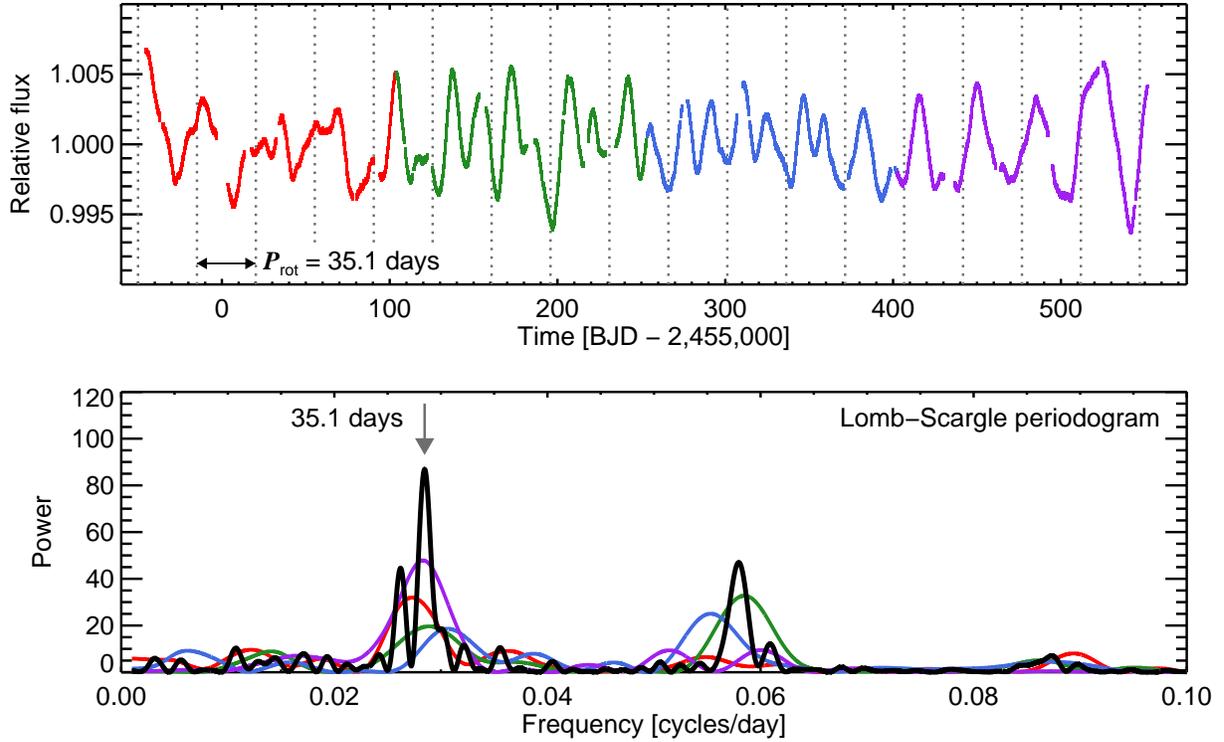}}
 \end{center}
 \vspace{-0.2in}


 \caption{ {\bf {\it Kepler} photometry.} 
   {\it Top.}---Time series of relative flux. Eclipse data have been
   removed. Vertical dotted
   lines are plotted every 35.1~days, the estimated rotation period.
   Different colors indicate the 4 segments for which periodograms
   were computed separately in order to gauge the uncertainty in the rotation period.
   {\it Bottom.}---Lomb-Scargle periodograms of the entire time
   series (thick black line) and each of the 4 segments
   (thinner colored lines).
   \label{fig:phot}}

\end{figure*}

\section{The Rossiter-McLaughlin effect}
\label{sec:rm}

We measured the sky-projected obliquity and rotation rate of the
primary star by conducting spectroscopic observations of a primary
eclipse and analyzing the Rossiter-McLaughlin (RM) effect. The RM
effect is the anomalous Doppler shift that is observed during eclipses
as a consequence of the selective blockage of the rotating stellar
photosphere (Rossiter 1924, McLaughlin 1924).

We used the Keck~I 10m telescope and HIRES spectrograph to gather 14
spectra on 2011~May~28/29, starting 40 minutes before ingress and
extending for 5~hours until morning twilight, thereby covering about
three-quarters of the eclipse. Another 3 spectra were obtained the
following night to track the out-of-eclipse velocity variation. The
typical exposure time was 19~minutes. The I$_2$ absorption cell was
used to establish the wavelength scale and instrumental profile. A
single exposure without I$_2$ was also obtained to serve as a template
spectrum. The relative radial velocities (RVs) were determined with a
descendant of the algorithm of Butler et al.~(1996). They are given in
Table~1 and plotted in Figure~2.

\begin{figure*}[ht]

\begin{center}
 \leavevmode
 \hbox{
 \epsfxsize=6.5in
 \epsffile{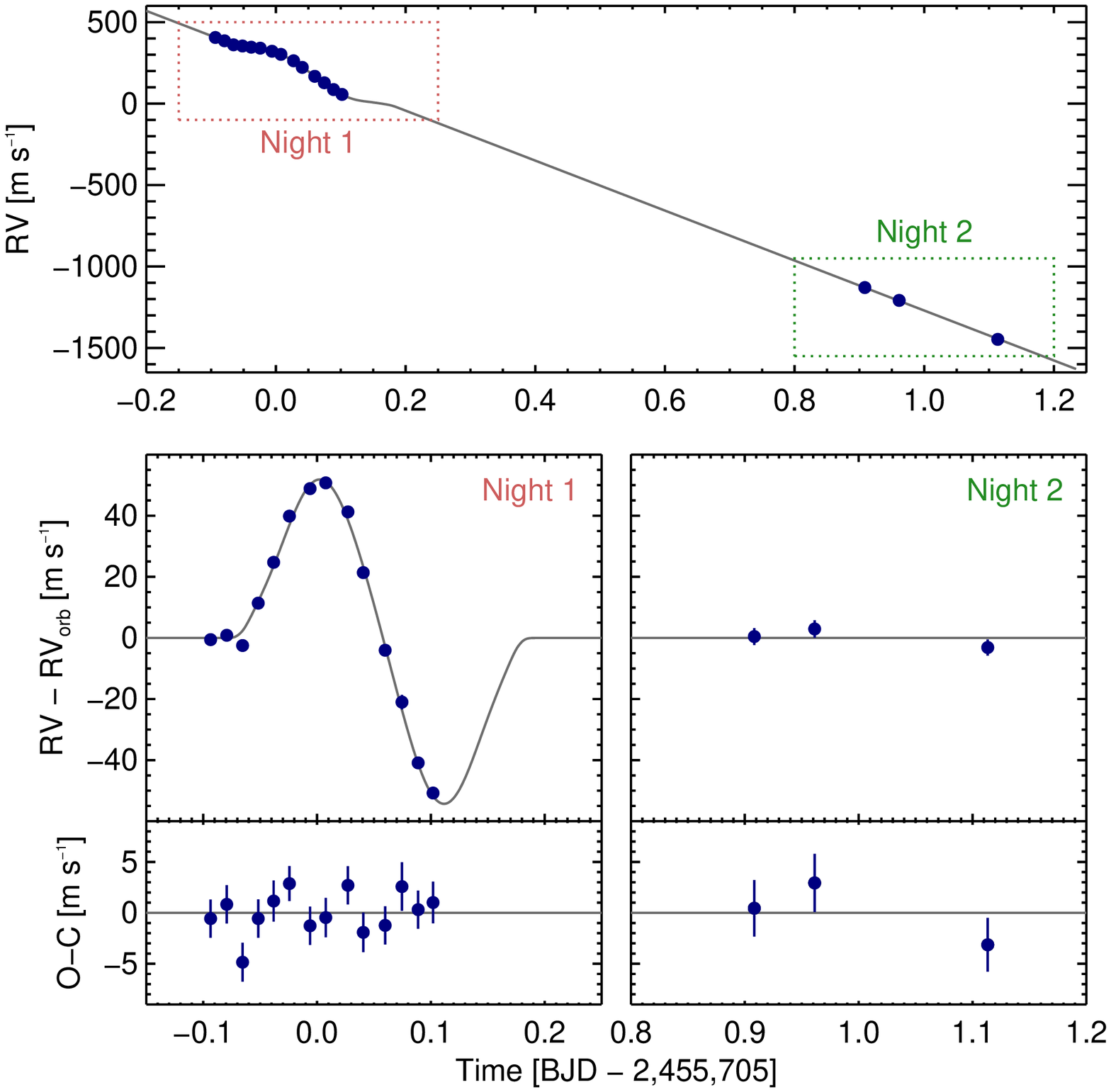}}
 \end{center}
 \vspace{-0.2in}


 \caption{ {\bf Keck radial velocities.} 
   {\it Top.}---Apparent radial velocity (solid points) and the
   best-fitting model (gray curve). {\it Bottom.}---After subtracting
   the best-fitting orbital model, thereby isolating the RM anomaly.
   Each night's data is shown separately, along with the residuals.
   \label{fig:rv}}

\end{figure*}

The ``red-then-blue'' pattern of the anomalous Doppler shift is
characteristic of a prograde orbit with good alignment between the
primary star's rotational and orbital angular momenta. In the first
half of the eclipse, the secondary covers the approaching (blue) half
of the primary, causing the net starlight to be redshifted; then, the
secondary moves over the receding (redshifted) half of the primary,
producing an anomalous blueshift. For quantitative modeling we used
the technique of Albrecht et al.~(2007), in which a pixellated stellar
disk is constructed, and a theoretical spectral line profile is
computed for each pixel based on the local intensity and velocity of
the photosphere. The integrated spectrum is obtained by summing over
the uneclipsed pixels, and then the RV is calculated by
cross-correlation with the integrated spectrum.

To compute the relative intensities of the pixels, we assumed a linear
limb-darkening law. The pixel velocities included the effects of
uniform rotation, macroturbulence, and the convective blueshift. The
model for macroturbulence was taken from Gray~(2005), assuming equal
radial and tangential velocity perturbations with a standard deviation
$\zeta_{\rm RT}=2.5$~km~s$^{-1}$. The model for the convective
blueshift was taken from Shporer \& Brown (2011), in which the
velocity of each pixel is shifted by $V_{\rm CB} = 0.2$~km~s$^{-1}$
away from the center of the star. The stellar radii and orbital
inclination were held fixed at the values determined by Doyle et
al.~(2011). We neglected any light from the secondary, as the light
ratio is constrained to be $<$1.3\% (Doyle et al.~2011).

The 6 adjustable parameters were the projected rotation rate $v\sin
i$, the projected obliquity $\beta$, the limb darkening coefficient
$u$, the central eclipse time $T_c$, the zero point of the relative RV
scale $\gamma$, and a parameter describing the out-of-eclipse RV
variation. For the latter, a simple linear slope would have sufficed,
but for convenience we used $M_B'$, the mass of the secondary when all
other orbital parameters (period, inclination, eccentricity, argument
of pericenter) are held fixed at the best-fitting values reported by
Doyle et al.~(2011). We expect small differences between $M_B'$ and
the true secondary mass $M_B$ due to the small uncertainties in the
other orbital parameters as well as spurious radial accelerations
caused by starspots or light from the secondary.

To determine the allowed parameter ranges we used a Markov Chain Monte
Carlo algorithm, with the the Metropolis-Hastings algorithm and Gibbs
sampler. The likelihood was taken to be $\exp(-\chi^2/2)$, where
$\chi^2$ is the usual sum of the standardized residuals between the
observed and calculated RVs. Uniform priors were adopted for $v\sin
i$, $\beta$, $\gamma$, and $M_2'$. A Gaussian prior was used for
$T_c$, with a central value equal to the predicted ephemeris time and
a standard deviation of 1~min (the typical level of eclipse timing
variations). A Gaussian prior was used for the limb darkening
coefficient, $u=0.8\pm 0.1$, based on the tables of Claret~(2000).

The formal ``1$\sigma$'' uncertainty interval was taken to be the
range between the 15.8\% and 84.2\% levels of the cumulative
distribution of the marginalized posterior for each parameter. In
addition, we checked on the sensitivity of the results to the assumed
values of the macroturbulent velocity $\zeta_{\rm RT}$ and convective
blueshift $V_{\rm CB}$, by perturbing each of those quantities by
$\pm$50\% and refitting. (It would be better to allow these parameters
to vary during the fit, but in practice this was too computationally
demanding.) Changing the macroturbulence had no appreciable
effect. Changing the convective blueshift caused shifts of
18~m~s$^{-1}$ in $v\sin i$ and $2^\circ$ in $\beta$, comparable to the
formal 1$\sigma$ intervals. Consequently, we enlarged the uncertainty
intervals for those two parameters by adding those shifts in
quadrature with the formal errors. Table~2 gives the results for all
the parameters.

\section{ Discussion }
\label{sec:discussion}

\subsection{Spin-orbit alignment}

The angle $\beta$ between the sky projections of the primary's
rotational angular momentum and the orbital angular momentum of the
binary was found to be $1.6 \pm 2.4$~degrees.\footnote{The angle
  $\beta$ is defined using the coordinate system of Hosokawa (1953). Note that Ohta et al.~(2005) and others have used a
  different symbol $\lambda$ to represent this angle, and a coordinate
  system such that $\lambda=-\beta$.} The star is apparently aligned
with the orbit to within a few degrees.

Ordinarily a warning must be issued here: the stellar rotation axis
might be inclined along the line of sight, and the sky-projected
obliquity $\beta$ might not be representative of the true
three-dimensional obliquity $\psi$. In this case, though, an upper
bound on the true obliquity is enforced by the combination $v\sin i =
0.920 \pm 0.025$~km~s$^{-1}$ from the RM analysis, $P_{\rm rot} =
35.1\pm 1.0$~d from the {\it Kepler} data, and $R_\star = 0.6489\pm
0.0013$~$R_\odot$ from the photometric-dynamical model of Doyle et
al.~(2011). Assuming $v = 2\pi R_\star / P_{\rm rot}$, and adopting a
isotropic prior for $i$ (uniform in $\cos i$), we find $\sin i =
0.994^{+0.006}_{-0.043}$ and $i=90^\circ \pm 9^\circ$. With 95.4\%
confidence, the true obliquity is $\psi < 18\fdg3$.

This system's angular momentum has 5 contributions---the stellar
orbit, the planetary orbit, the primary rotation, the secondary
rotation, and the planetary rotation---with magnitudes in the
approximate ratios $10000 : 40 : 1 : 0.1 : 0.001$.  Kepler-16 is an
orderly system, with good alignment between the three largest portions
of the angular momentum. To our knowledge, Kepler-16 is the
longest-period stellar binary for which a stellar obliquity has been
measured (see Table~1 of Albrecht et al.~2011).

\subsection{Activity, rotation, and age}

The Keck spectra show Ca~H\&K chromospheric emission with $\log
R'_{\rm HK} = -4.68\pm 0.10$, stronger than the Sun's value of
$-4.91$. The spectra obtained previously by Doyle et al.~(2011) also
exhibit Ca~H\&K emission, with RV variations tracking those of the
primary, proving that the emission originates on the primary rather
than the secondary or the planet. We may therefore use empirical
relations between the rotation, chromospheric activity, and age of
main-sequence dwarfs to estimate the age of Kepler-16~A and check
whether there is anything unusual about its properties.

Building on work by Barnes~(2007) and others, Mamajek \&
Hillenbrand~(2008) provided up-to-date activity/rotation/age relations
for K2-F7 dwarfs ($\approx$0.8-1.3~$M_\odot$), which we extrapolated
to interpret the 0.7~$M_\odot$ primary of Kepler-16. The activity/age
relation in their Eqn.~(3) gives an age of $1.8\pm 1.2$~Gyr. The
rotation/age relation implicit in their Eqns.~(12-14) give an age of
$3.7\pm 0.8$~Gyr.  A similar age is obtained from the relations of
Barnes~(2010) and Barnes \& Kim~(2010).

In addition, the rotation and activity are expected to be directly
linked, with a particularly strong correlation betwen the Rossby
number Ro (the ratio of rotation period to convective turnover
timescale) and $\log R'_{\rm HK}$. We used Eqn.~(4) of Noyes et
al.~(1984) to estimate the convective turnover timescale, obtaining
$\tau_c = 23.0$~d and Ro~$=1.52$. The relation between Ro and
chromospheric activity shown in Fig.~7 of Mamajek \& Hillenbrand
(2008) predicts $\log R'_{\rm HK}=-4.70$, in good agreement with the
measured value. The light variations of 0.5-1\% are also typical for a
star with this level of chromospheric emission (see, e.g., Hall et
al.~2009).

All together, the rotation period, chromospheric emission level, and
fractional light variations paint a picture of an ordinary
0.7~$M_\odot$ dwarf star with an age of 2-4~Gyr.

\subsection{Comparison to evolutionary models}

The Keck spectrum provides new estimates of the primary's photospheric
parameters. Analysis with Spectroscopy Made Easy, a software package
written by Valenti \& Piskunov (1996), gives $T_{\rm eff} = 4337\pm
80$~K and [Fe/H]~$= -0.04\pm 0.08$ for a fixed value of $\log g =
4.6527$ (the value determined by Doyle et al.~2011, which has a
negligible uncertainty for this purpose). These agree with the
previously reported results $T_{\rm eff} = 4450\pm 150$~K and
[m/H]~$=-0.3\pm 0.2$ (Doyle et al.~2011).

Figure~3 shows a comparison between the observed masses and radii of
Kepler-16~A and B and the theoretical evolutionary models of Baraffe
et al.~(1998). Models are shown for 1 and 5~Gyr for a mixing-length
parameter of $1.0$. The 1~Gyr model gives the best match to the
primary mass and radius, although it is also possible to obtain a good
fit at 5~Gyr by increasing the mixing-length parameter toward the
Solar value of 1.9 (as might be expected for the primary; see, e.g.,
Demory et al.~2009). The models also match the measured effective
temperature of the primary. Interestingly the calculated primary
radius is within a few percent of the observed radius. This is in
contrast to the other stars of similar mass shown in Figure~3, for
which the observed radii are 10-15\% larger than the model
predictions. Such discrepancies have been attributed to high activity
and rapid rotation in the stars that have been studied closely (see,
e.g., Chabrier et al.~2007). The secondary star's mass and radius are
also within a few percent of the calculated values. We leave a more
detailed comparison with models for future work, which should take
into account the uncertainty in the metallicity as well as the
$\alpha$-element abundances (which have not yet been measured).

\begin{figure*}[ht]

 \begin{center}
  \leavevmode
\hbox{
  \epsfxsize=5.0in
  \epsffile{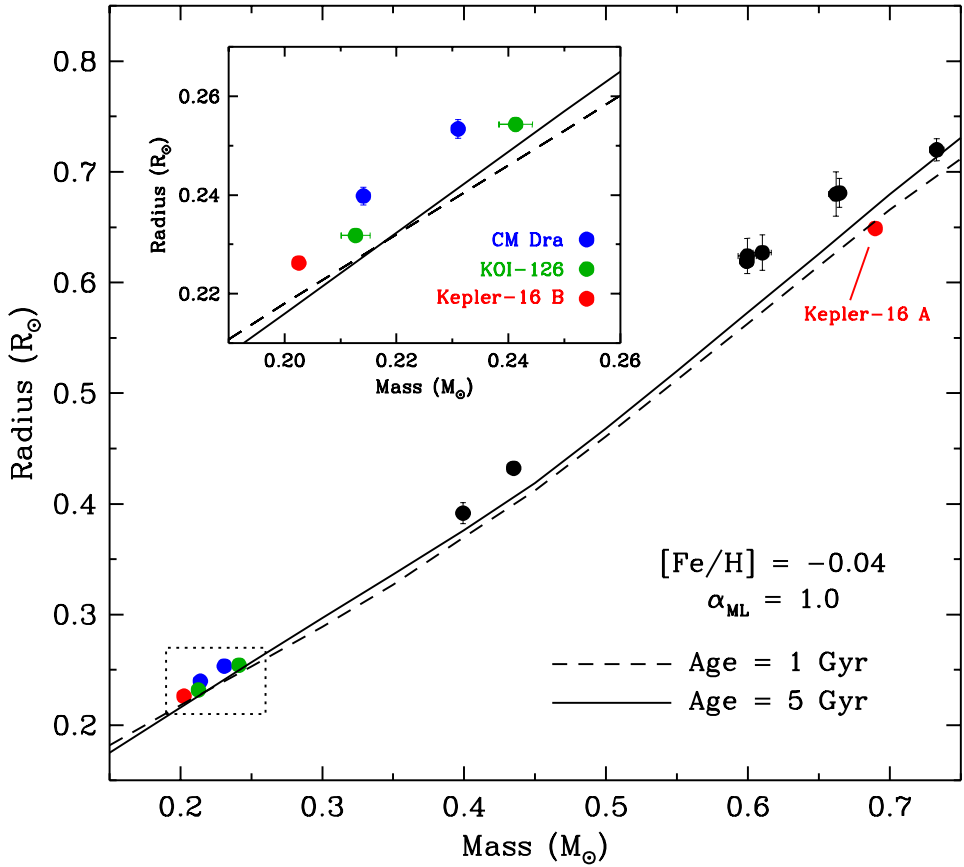}}
 \end{center}
 \vspace{-0.1 in}


\caption{{\bf Theoretical and observed masses and radii of low-mass
    stars.} The model curves are from Baraffe et al.~(1998),
    for a metallicity of $-0.04$ (obtained by linear interpolation
    between $0.00$ and $-0.50$) and a mixing-length parameter
    $\alpha_{\rm ML} = 1.0$.
    In addition to Kepler-16 (Doyle et al.~2011), data are shown
    for other low-mass stars for which the masses and radii have
    been determined to within 3\% according to the rigorous
    criteria used by Torres et al.~(2010). Those systems are:
    KOI-126 (Carter et al.~2011),
    YY~Gem (Torres \& Ribas 2002),
    CU~Cnc (Ribas 2003),
    GU~Boo (Lopez-Morales \& Ribas 2005),
    NGC~2204~S892 (Rozyczka et al.~2009),
    and IM~Vir (Morales et al.~2009).
    \label{fig:mr}}

 \vspace{0.3 in}
\end{figure*}

\subsection{Formation scenarios}

Binary stars are thought to form by fragmentation of collapsing
prestellar cores, with close binaries such as Kepler-16 possibly
resulting from fragmentation during a late, isothermal phase of the
collapse (see, e.g., Goodwin et al.~2007). Stable hierarchial triples
can also be produced by fragmentation (Boss 1991, Bate 2009), raising
the question of whether the three bodies of Kepler-16 formed in this
manner.

In this scenario one would expect a tendency for the fragments to be
aligned, but there is no obvious reason why the alignment would be as
close as is observed for Kepler-16. Among the triple systems that
formed in a large-scale simulation by Bate~(2009), the orbital planes
were typically misaligned by $\approx$60$^\circ$. Furthermore there
are some close binaries with much larger obliquities than Kepler-16~A,
such as DI Herculis (Albrecht et al.~2009).

This suggests an additional chapter is needed in the story, beyond
fragmentation. Perhaps after their formation the stars continued to
accrete substantially from a circumbinary disk, which would have
decreased their orbital separation and aligned their spin axes (Bate
et al.\ 2002). Or the primary could have formed with a massive
circumstellar disk, which then fragmented to form the secondary. In
either of these scenarios, the planet could have formed by core
accretion near the inner edge of the circumbinary disk (see, e.g.,
Pierens \& Nelson 2008, Marzari et al.\ 2008).

\subsection{Evidence for tidal evolution}

Another process that could have reduced the stellar obliquity is tidal
evolution. In the fullness of time, tides synchronize and align the
spins of a binary system, and circularize their orbit (Zahn 1977, Hut
1981). Conventional wisdom would say that tides are irrelevant for
Kepler-16 due to the relatively long orbital period (41~days) and wide
separation ($a=0.22$~AU, $a/R_\star=74$), and indeed the orbital
eccentricity of 0.16 shows that circularization is incomplete. On the
other hand, synchronization and alignment should be faster than
circularization because the rotational angular momentum is smaller
than the orbital angular momentum. More broadly, tidal evolution
timescales are poorly known, especially for spin evolution (see, e.g.,
Mazeh 2008).

An order-of-magnitude assessment of the importance of tidal evolution
begins with the observation that most late-type binaries with periods
$<$10~days have circular orbits (Mazeh 2008). Therefore the
circularization timescale is $\sim$5~Gyr for $a/R_\star \sim 20$. If
this timescale varies as $(a/R_\star)^8$ (Zahn 1977), then for
Kepler-16 it is $5~(74/20)^8 \sim 2\times 10^5$~Gyr, i.e., consistent
with the observation that the orbit is still eccentric. We further
suppose that the spin evolution timescale is smaller by a factor of
$\sim$$10^4$ (the ratio of orbital to rotational angular momenta) and
varies as $(a/R_\star)^6$ (Zahn 1977). Then the spin evolution
timescale for Kepler-16 would be $5\times 10^{-4}$~$(74/20)^6
\sim$~1~Gyr. This is corroborated by the more detailed tidal model of
Terquem et al.~(1998): using their Eqns.~(41-42), the timescales for
circularization and synchronization are $10^4$~Gyr and 2~Gyr,
respectively. These calculations are subject to the well-known
uncertainties in tidal evolution timescales, but they do suggest that
the low obliquity of Kepler-16~A is at least partly a consequence of
tidal evolution.

In this context, the rotation period of the primary is
intriguing. According to the tidal theory of Hut (1981), before the
orbit circularizes the stellar spins evolve into a pseudosynchronous
state, with spin periods shorter than the orbital period due to the
enhanced tidal dissipation at pericenter passages. For Kepler-16~A,
using Eqn.~(42) of Hut~(1981) with $e=0.15944$ (Doyle et al.~2011),
the predicted pseudosynchronous period is $35.62$~days, in agreement
with the measured period of $35.1\pm 1.0$~days. As noted in
Section~4.1, though, the measured rotation period is not unusual even
for an isolated star, and the agreement with the pseudosynchronous
value may be a coincidence. A more detailed study of the tidal
evolution of this unique triple system is warranted, as are further
spin-orbit studies of relatively wide binaries.

\acknowledgments We thank Brice-Olivier Demory, Nevin Weinberg, and
Jamie Lloyd for helpful discussions. Work by J.N.W.\ and S.A.\ was
supported by NASA Origins award NNX09AB33G.  G.T.\ acknowledges
partial support from the NSF through grant AST-1007992. Funding for
the {\it Kepler} Discovery mission is provided by NASA's Science
Mission Directorate. The W.M.~Keck Observatory is operated as a
scientific partnership among the California Institute of Technology,
the University of California, and the National Aeronautics and Space
Administration, and was made possible by the generous financial
support of the W.M.~Keck Foundation. We extend special thanks to those
of Hawaiian ancestry on whose sacred mountain of Mauna Kea we are
privileged to be guests.

Facilities: \facility{Keck(HIRES)}, \facility{Kepler}.

\eject

\clearpage

\begin{deluxetable}{lcc}

\tabletypesize{\normalsize}
\tablewidth{0pt}
\tablecaption{Relative Radial Velocity Measurements of Kepler-16~A \label{tbl:rv}}

\tablehead{
\colhead{BJD$_{\rm UTC}$} &
\colhead{RV [m~s$^{-1}$]} &
\colhead{Unc.~[m~s$^{-1}$]}
}

\startdata
  $  2455704.90651$  &  $    405.81$  &  $   1.88$  \\
  $  2455704.92067$  &  $    385.56$  &  $   1.88$  \\
  $  2455704.93459$  &  $    360.90$  &  $   1.91$  \\
  $  2455704.94835$  &  $    353.70$  &  $   1.88$  \\
  $  2455704.96188$  &  $    346.34$  &  $   2.02$  \\
  $  2455704.97574$  &  $    340.26$  &  $   1.72$  \\
  $  2455704.99387$  &  $    321.48$  &  $   1.88$  \\
  $  2455705.00766$  &  $    302.26$  &  $   1.94$  \\
  $  2455705.02716$  &  $    262.90$  &  $   1.88$  \\
  $  2455705.04061$  &  $    222.41$  &  $   1.94$  \\
  $  2455705.05985$  &  $    167.53$  &  $   1.87$  \\
  $  2455705.07461$  &  $    127.95$  &  $   2.39$  \\
  $  2455705.08883$  &  $     86.27$  &  $   1.88$  \\
  $  2455705.10192$  &  $     56.34$  &  $   2.05$  \\
  $  2455705.90834$  &  $  -1129.19$  &  $   2.78$  \\
  $  2455705.96133$  &  $  -1207.97$  &  $   2.86$  \\
  $  2455706.11341$  &  $  -1447.21$  &  $   2.64$  
\enddata

\clearpage

\end{deluxetable}

\begin{table*}
  \begin{center}
    \centerline{Table 2: Summary of Results\label{tbl:results}}
    \begin{tabular}{l  r@{~$\pm$~}l}

      \tableline\tableline
      \noalign{\smallskip}
         Parameter & \multicolumn{2}{c}{~~~~~~~~~~~~Value} \\
      \noalign{\smallskip}
      \hline
      \noalign{\smallskip}
      Projected rotation speed, $v\sin i$~[km~s$^{-1}$]   & $0.920$ & $0.025$ \\  
      Projected spin-orbit angle, $\beta$~[deg]        & $1.6$ & $2.4$ \\ 
      Secondary mass parameter, $M_B'$~[$M_\odot$]        & $0.20133$ & $ 0.00026$ \\  
      Velocity zero point, $\gamma$~[m~s$^{-1}$]    & $427.6$ & $1.3$ \\
      Limb darkening coefficient, $u$                    & $0.717$ & $0.079$ \\
      Central eclipse time, $T_c$~[BJD$_{\rm UTC}$]         & 2,455,705.05388 & $0.00069$ \\
      \noalign{\smallskip}
     \hline
      \noalign{\smallskip}
      Stellar rotation period, $P_{\rm rot}$~[days]       & $35.1$ & $1.0$ \\
      Stellar inclination angle, $i$~[deg]              & $90$ & $9$ \\
      Stellar obliquity, $\psi$~[deg]                   & \multicolumn{2}{r}{$<$18.3 (95.4\% conf.)} \\
      \noalign{\smallskip}
     \hline
      \noalign{\smallskip}
      Effective temperature, $T_{\rm eff}$~[K]            & $4337$ & $80$ \\
      Iron abundance, [Fe/H]                             & $-0.04$ & $0.08$ \\
      Surface gravity, $\log (g$~[cm~s$^{-2}$)]          & $4.6527$ & $0.0017$ \\
      Main-sequence age [Gyr]                            & $3.0$ & $1.0$ \\
      \noalign{\smallskip}
      \tableline
      \noalign{\smallskip}
      \noalign{\smallskip}
    \end{tabular}

    \tablecomments{All the stellar parameters refer to the primary
      star, Kepler-16~A, except for the secondary mass parameter
      $M_B'$ (defined in the text). The result for $M_B'$ differs by
      $0.00120\pm 0.00071$~$M_\odot$ from the secondary mass
      $M_B=0.20255\pm 0.00066$~$M_\odot$ determined by Doyle et
      al.~(2011).}

  \end{center}
  
\end{table*}

\end{document}